\documentclass[fleqn,twoside]{article}
\usepackage{espcrc2} 
\usepackage{graphicx}

\newcommand{\AmS}{{\protect\the\textfont2
A\kern-.1667em\lower.5ex\hbox{M}\kern-.125emS}}

\title{Neutrino Interactions at Low and Medium Energies}
\author{E.A.~Paschos\address{Institut f\"{u}r Physik, Universit\"{a}t 
Dortmund\\D-44221 Dortmund, Germany}%
        \thanks{Invited Talk presented at the Workshop on 
 Low Energy Neutrino--Nucleus Interactions, 
Dec.\ $13-16$, 2001, KEK, Tsukuba, Japan.}}
       
\begin{document}

\begin{abstract}
We discuss the calculations for several neutrino 
induced reactions from low energies to the GeV region.
Special attention is paid to nuclear corrections when the targets 
are medium or heavy nuclei.
Finally, we present several ratios of neutral to charged current 
reactions whose values on isoscalar targets can be estimated 
accurately. The ratios are useful for investigating 
neutrino oscillations in Long Baseline experiments.

\end{abstract}

\maketitle

\section{Introduction}

As I mentioned in my previous talk there are many calculations
available for low and intermediate energy neutrino 
interactions. Some of them have been compared with
experimental data providing tests for theoretical models 
(parton model, neutral currents, ...) and 
were useful in extracting parameters (Weinberg angle, $V_{cs}$, ...).
Other processes were calculated at the level of our theoretical understanding
and are still waiting to be compared with data.
In some cases, the comparisons
are necessary for the correct interpretation of the results.

The topics that I will cover are the following:
\begin{enumerate}
\item[1)] Cross sections produced by $\nu_{\mu}$ and
$\nu_{\tau}$ neutrinos, including threshold effects for
the $\tau$--lepton.
\item[2)] The main reactions to be discussed are 
\begin{itemize}
\item[--]{Deep inelastic scattering}
\item[--]{Resonance production\\
$\quad\quad I=3/2$ ($\Delta$--resonance)\\
$\quad\quad I=1/2$ ($P_{11},S_{11}$ resonances)
\item[--]{Quasi--elastic Scattering}}
\end{itemize}
\item[3)] I will cover cross sections on free protons
and loosely--bound neutrons, as well as reactions with
medium--heavy nuclei as targets. The nuclei bring in additional
corrections whose origin is different for each of the
above reactions.
\item[4)] At the end, I will mention the ionization
of atoms by neutrinos with energies in the keV range.
\end{enumerate}
My recent work was done in collaboration with 
L.\ Pasquali and J.Y.\ Yu \cite{eap,yuji}, where you can find
more details.

\section{Deep Inelastic Scattering (DIS)}

The general structure of the differential cross section
is well known. New terms appear when we keep 
the masses of heavy
leptons which are important for reactions 
induced by $\tau$--neutrinos.  
The hadronic tensor includes now two additional structure functions
\begin{eqnarray}
W_{\mu\nu}
& = &
- g_{\mu\nu} F_1(x,Q^2) + \frac{p_{1\mu}p_{1\nu}}{{p_1\cdot q}} F_  
            2(x,Q^2) \nonumber \\
&-& i \epsilon_{\mu\nu\rho\sigma} \frac{p_1^{\rho}
            q^{\sigma}}{2 p_1\cdot q} F_3(x,Q^2)\nonumber \\
&+& \frac{q_{\mu} q_{\nu}}{p_1\cdot q} F_4(x,Q^2) \nonumber \\
&+& (p_{1\mu}q_{\nu} + p_{1\nu}q_{\mu}) F_5(x,Q^2).\nonumber
\end{eqnarray}
with $p_1$ and $q$ the momenta of the target nucleon and the exchanged current,
respectively.
We use the standard notation with 
$x = \frac{Q^2}{2 M\nu}, \, Q^2 = -q^2$, $y=\frac{\nu}{E_\nu}$ and 
$\epsilon_{\mu\nu\rho\sigma}$ the antisymmetric tensor.

The differential cross section has the general form
\begin{eqnarray}
\frac{{\rm d}\sigma^{\nu,\bar{\nu}}}{{\rm d}x{\rm d}y}
& = &
     \frac{G_F^2 M_N E_\nu}{\pi} 
     \bigg [y\Big(xy + \frac{m_l^2}{2 E_\nu M_N}\Big)F_1  \nonumber\\
& + & 
    \Big(1-y -\frac{M_Nxy}{2 E_\nu} - 
     \frac{m_l^2}{4 E_\nu^2}\Big) F_2 \nonumber\\
&\pm & 
\Big(xy(1-\frac{y}{2})-y\frac{m_l^2}{4 M_N E_\nu}\Big) F_3\nonumber\\ 
& + & 
\Big(x y \frac{m_l^2}{2 M_N E_\nu} 
+ \frac{m_l^4}{4 M_N^2 E_\nu^2}\Big) F_4 \nonumber\\
& - &
\frac{m_l^2}{2 M_N E_\nu} F_5\bigg],\nonumber
\end{eqnarray}
\\
$x = \frac{Q^2}{2 M_N \nu}$ with $\nu = E_\nu-E_l$, $Q^2 = 2 M_N E_\nu x y$, 
and $M_N$ with $N = p,n$ the nucleon mass. 
The two signs $\pm F_3(x)$ correspond to $\nu$-- and
$\bar{\nu}$--nucleon scattering.  For the structure
functions we use the relations
\begin{eqnarray*}
2 x F_1 & = & F_2\quad\quad
 {\rm{Callan-}}{\rm{Gross}}\,\,{\rm{relation}}\\
F_4 & = & 0\quad\quad\,\,
  {\rm{Albright-}}{\rm{Jarlskog}}\,\,{\rm{relation}} \,\cite{albright}\\
x F_5 & = & F_2\quad\quad{\rm{spin}}\,\, \frac{1}{2}\,\,
   {\rm{constituents}}
\end{eqnarray*}
which follow from the scattering on free quarks.  The
structure functions $F_2,\, F_3$ in terms of quark
distributions are known.  For $\nu_{\mu}$ reactions we
use structure functions above the charm threshold and
for $\nu_{\tau}$ those below the charm threshold.  
\begin{figure}[htb]
\vspace{-1.2cm}
\hspace{-1.cm}
\includegraphics[angle = 0,width =18pc,height = 15pc]{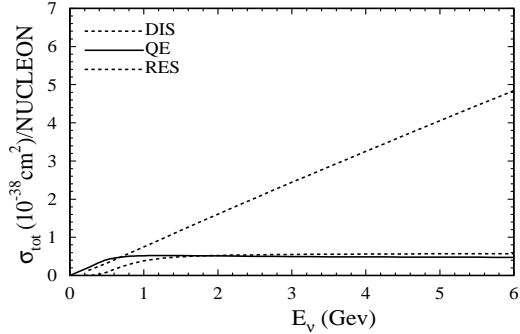}
\vspace{-2.cm}
\caption{The total cross sections of DIS, QE and RES}
\label{fig1}
\end{figure}
\vspace{-0.6cm}

As a warm up, I show in Fig.~(\ref{fig1}) the total cross section at low and
medium energies as the incoherent sum of quasi--elastic,
resonance production and deep inelastic scattering. At very low 
energies $E_{\nu}< 1.0$ GeV the quasi--elastic reaction (QE)
dominates, then opens the channel for $\Delta$--production
and other resonances (RES) 
and finally appear multi--particle final states (DIS).  The
dynamics for each process and the corrections are
different and we shall discuss them separately. 
The contributions of quasi-elastic scattering
and resonances are clearly evident in the data up to $\sim$ 2 GeV.
They produce a peculiar structure, looking like a step.
Above this energy the charged current cross sections are known to rise
linearly with energy.  

\begin{figure}[htb]
\vspace{-1.2cm}
\hspace{-1.cm}
\includegraphics[angle = 0,width =18pc,height = 15pc]{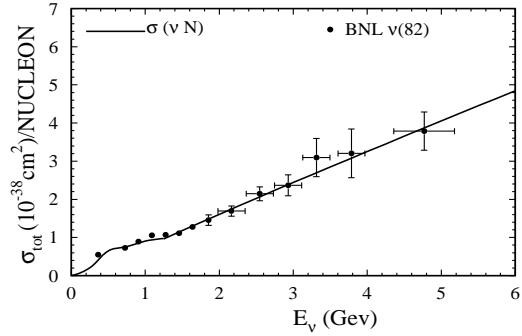}
\vspace{-2.cm}
\caption{The total cross section of RES+QR+DIS}
\label{fig2}
\end{figure}
\vspace{-0.3cm}

Fig.~(\ref{fig2}) shows the total cross
sections for $E_{\nu}< 6$ GeV with the data from Ref \cite{baker}. 
Above $2\, {\rm GeV}$ there is 
a linear rise of the cross section with $E_\nu$.
Their slopes have been
measured precisely up to $300\, {\rm GeV}$.

\begin{figure}[htb]
\vspace{-0.7cm}
\hspace{-0.8cm}
\includegraphics[angle = 0,width =18pc,height = 15pc]{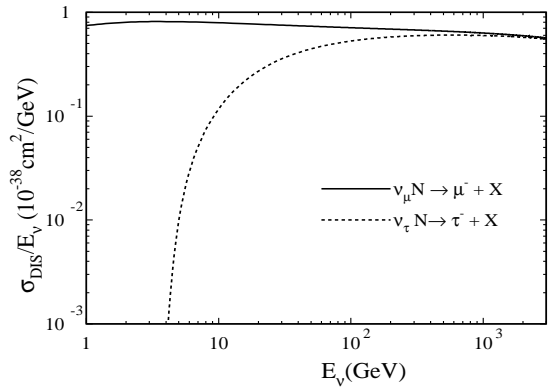}
\vspace{-1.cm}
\hspace{-1.2cm}
\caption{The total cross section for $\nu_{\mu}$'s and $\nu_{\tau}$'s}
\label{fig3a}
\end{figure}

In addition, I show in Fig.~(\ref{fig3a})
the slopes of the total cross section
for $\nu_{\mu}$'s and $\nu_{\tau}$'s, where the threshold dependence is
now evident.

\subsection{Nuclear Corrections to DIS}
\vspace{0.4cm}
Neutrino reactions have also been measured in nuclei
where the need for corrections became evident.  For
deep inelastic scattering nuclear corrections are
incorporated by modifying the parton distribution
functions. We use two sets of distribution functions.
\begin{enumerate}
\item[1)] The $\chi^2$ analysis of Hirai, Kumano and
Mijama \cite{hkm}, and
\item[2)] the evolution analysis of Eskola, Kokhinen
and Solgado \cite{eks}.
\end{enumerate}
The analyses include shadowing, anti--shadowing, EMC--effect and Fermi-motion
effects.  The same parton distributions were used for
$\nu_{\mu}$ and $\nu_{\tau}$ reactions.

\begin{figure}[htb]
\vspace{-1.5cm}
\hspace{-0.8cm}
\includegraphics[angle = 0,width =18pc,height = 15pc]{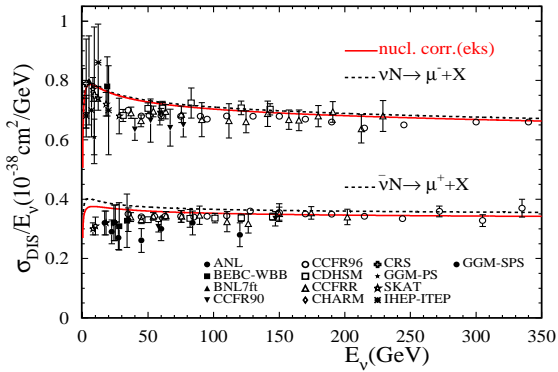}
\vspace{-1.cm}
\hspace{-1.2cm}
\caption{The total cross section of DIS for $\nu_{\mu}$'s and $\nu_{\tau}$'s}
\label{fig3b1}
\end{figure}
\vspace{-0.4cm}

Fig.~(\ref{fig3b1}) compares the experimental slopes 
with theoretical calculation.
The dashed curve is without nuclear corrections 
and the solid curve includes nuclear corrections \cite{ref7}.
The calculated cross sections and the data agree with
each other.  For the  total cross sections the nuclear
correlations are small as shown in Figure~(\ref{fig3b1}).

We mention briefly that particle
multiplicities and their momentum distributions can
be measured for each structure function separately.
Subsequently, they can be used to calculate 
multiplicities and their properties in $\nu_{\tau}$
induced reactions.

\section{Resonance Production}

The production of the $\Delta$--resonance has been
calculated in terms of the form factors.  Formulas
are available for triple differential decays
\begin{eqnarray}
\frac{{\rm d}\sigma}{{\rm d}Q^2{\rm d} W{\rm d}\cos\theta_\pi} 
&=& \frac{G_F^2}{16\pi M_N^2}\sum_{i=1}^{3}\big[K_i\widetilde{W}_i \nonumber\\
&-& 
\frac{1}{2} K_i D_i (3 \cos^2\theta_\pi - 1)\big],\nonumber
\end{eqnarray}
as well as integrated cross--sections.  We considered two models,
one for $\Delta$--production and another one for the
excitation of the $I=1/2$ resonances:  $P_{11},\, S_{11}$.
The $K_i$'s are kinematic factors and the $W_i$'s 
contain the hadronic structure in terms of the form factors.
For the $\Delta$--resonance the following form factors were used
\begin{eqnarray}
C_3^V(Q^2) &=& \frac{2.05}{(1+\frac{Q^2}{0.54 \, {\rm GeV}^2})^2}\nonumber\\
C_4^V(Q^2) &=& -\frac{M_N}{M_\Delta} C_3^V\nonumber\\
C_5^V(Q^2) &=& 0,\nonumber\\
C_k^A (Q^2) &=& C_k (0) \Big(1 + \frac{a_k Q^2}{b_k+Q^2}\Big) 
\Big(1 + \frac{Q^2}{M_A^2}\Big)^{-2}\nonumber\\
C_6^A(Q^2) &=& 
      \frac{g_\Delta f_\pi}{2\sqrt{3} M_N}\frac{M^2}{m_\pi^2 + Q^2}\nonumber
\end{eqnarray}
with $k = 3,4,5, \, C_3^A(0) = 0, \, C_4^A(0) = - 0.3, \, 
C_5^A(0) = 1.2$, $a_4 = a_5 = -1.21, \, b_4 = b_5 = 2 \,  {\rm GeV^2}$ 
and the axial mass $M_A = 1.0 \,{\rm GeV}$. 
The rest of the
formalism can be found in Ref.~\cite{schreiner}.

For the charged current cross--section there are data
for three channels and we show the comparison with 
our calculation. In Figures~(\ref{fig4})--(\ref{fig6}), 
the agreement at low energies is at
the 10$\%$ level.  The experimental points have substantial
errors which should be improved.

\subsection{Production of Nuclear Targets}
\vspace{0.5cm}
The Long Baseline (LBL) experiments measure reactions on nuclear targets,
like $_8{\rm{O}}^{16},\, _{18}{\rm{Ar}}^{40}$ and 
$_{26}{\rm{Fe}}^{56}$. 
To a first approximation the cross--section is the incoherent
sum of protons and neutrons. Nuclear effects are
now relevant.  The neutrinos scatter throughout the 
nucleus with a probability proportional to the nuclear
density, which is considered to be isoscalar (the
average cross--section on neutrons and protons).  The
neutrino--nucleon cross--section is affected 
\begin{enumerate}
\item[i)] by the Fermi--motion of the nuclei, which 
shifts the energy and
\item[ii)] by the Pauli--suppression (Pauli--blocking), i.e. when a
nucleon receives a small momentum transfer, below the
Fermi--sea, it cannot recoil because the state to 
which it must go is already occupied.
\end{enumerate}
These two effects are frequently taken into account by using nuclear
parameters.  The Pauli suppression, for instance, is
3-5$\%$ for $_{26}Fe^{56}$.

When the pion is observed there is a third effect.  As
the pions wander through the nucleus, they scatter
performing a random walk.
We view the scattering as a two step process \cite{anp}.\\

\noindent{\bf\underline{step 1}}: 
The reaction
\begin{eqnarray}
\nu +\frac{(n+p)}{2}\rightarrow l+\frac{(n'+p')}{2} +\pi^{\pm,0} \nonumber
\end{eqnarray}
takes places in the nucleus being, proportional to the
density profile of the nucleus.
For the charge density for $_8{\rm{O}}^{16}$ we use 
the Harmonic Oscillator Model
and for $_{18}{\rm{Ar}}^{40}$, $_{26}{\rm{Fe}}^{56}$
the Fermi model.
We denote this original production cross sections by 
${\rm d}\sigma (N_T;+),\,
{\rm d}\sigma (N_T;0),{\rm d}\sigma (N_T;-)$.

\noindent{\bf\underline{step 2}}: 
The subsequent random walk process is
analyzed as a multiple scattering of $\pi$'s from
nucleons (protons and neutrons) using the well known
low energy pion--nucleon cross--sections. 
At each scattering the pions may retain or exchange 
their charge, and the original population of pion changes.
The phenomenon is described by a transition matrix $M$ as follows
\begin{eqnarray}
{\left(\begin{array}{c}\displaystyle
{{\rm d}\sigma(_ZT^A;+)}\\
\displaystyle{{\rm d}\sigma(_ZT^A;0)}\\
\displaystyle{{\rm d}\sigma(_ZT^A;-)}
\end{array}\right)} &=& 
M {\left(\begin{array}{c}\displaystyle
{{\rm d}\sigma(N_T;+)}\\
\displaystyle{{\rm d}\sigma(N_T;0)}\\
\displaystyle{{\rm d}\sigma(N_T;-)}
\end{array}\right)}. \nonumber
\end{eqnarray}
The transition matrix has a simpler form dictated by isospin conservation
\begin{eqnarray}
M = A \left(\begin{array}{ccc}
{1-c-d} & d & c \\
d & {1-2d} & d \\
c & d & {1-c-d}
\end{array}\right) \nonumber
\end{eqnarray}
with $c$ and $d$ being determined by the random walk process \cite{anp}. 

\begin{figure}[htb]
\vspace{-1.4cm}
\hspace{-2.5cm}
\includegraphics[angle = 0,width =30pc,height = 30pc]{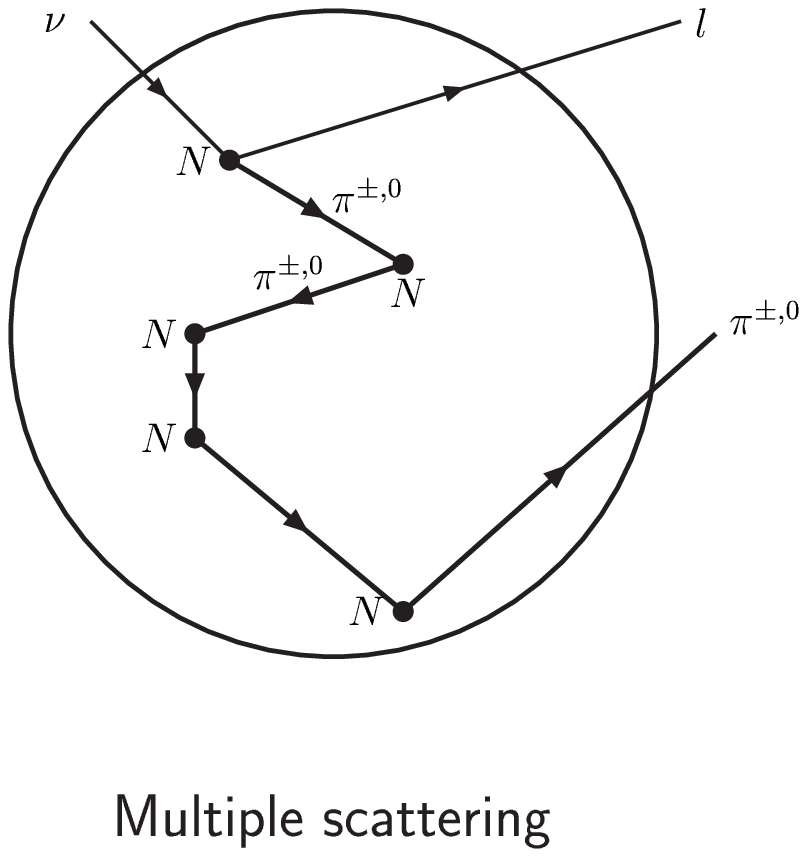}
\end{figure}
\vspace{-7.cm}

The overall factor A contains the 
Pauli suppression factor
$g(w,Q^2)$ and the transport function $f(\lambda)$ with the eigenvalue 
$\lambda = 1,\, 1/2$ and $5/6$.
This is a model developed some time ago \cite{anp}
for the interpretation of neutral current events (ANP model).
It has been tested in a few cases.  

In the ANP model the transition matrix on iron is \cite{eap,slad}
\begin{eqnarray}
M(_{26}Fe^{56}) = A \left(\begin{array}{ccc}
0.72 & 0.19 & 0.09 \\
0.19 & 0.61 & 0.19 \\
0.09 & 0.19 & 0.72
\end{array}\right),\nonumber 
\end{eqnarray}
where $A = g(W,Q^2) f(\lambda=1)$ with $g(M_\Delta, Q^2)\approx 0.96$
and $f(\lambda =1)= 0.625$.
I will give two examples to demonstrate how the corrections work.\\

\noindent{\bf\underline{Example 1}}:
We consider a neutrino beam of $E_{\nu}=1.5$ GeV.
We take the neutral current cross sections
\begin{eqnarray}
\sigma(\nu p\rightarrow \nu n \pi^+) 
                    &=&\sigma(\nu n\rightarrow \nu p \pi^-) \nonumber\\
                    &=& 0.064\times 10^{-38} {\rm cm^2} \nonumber\\
\sigma(\nu p\rightarrow \nu p \pi^0) 
                   &=&\sigma(\nu n\rightarrow \nu n \pi^0)\nonumber\\ 
                   &=& 0.125\times 10^{-38} {\rm cm^2} \nonumber
\end{eqnarray}
The relations between neutron and proton cross section follow from charge 
symmetry.
Applying the formalism described in step 2, I obtain
\begin{eqnarray}
2\times d\sigma(T;\pi^+) &=& [A (0.72+0.09)\times 0.064\nonumber\\
               &+& A (0.19\times 0.125)]\times 10^{-38}{\rm cm^2}\nonumber\\
               &=& A (0.052+0.024)\times 10^{-38}{\rm cm^2} \nonumber\\
               &=& A\times 0.075\times 10^{-38}{\rm cm^2}\nonumber
\end{eqnarray}
and similarly
\begin{eqnarray}
2\times d\sigma(T;\pi^0) &=& [A(0.19\times 2\times 0.064)\nonumber\\
               &+& A(0.62\times 0.125)]\times 10^{-38}{\rm cm^2} \nonumber\\
               &=& A (0.024+0.078)\times 10^{-38} {\rm cm^2} \nonumber\\
               &=& A\times 0.10\times 10^{-38} {\rm cm^2}.\nonumber
\end{eqnarray}
We notice that, beyond the overall factor $A$, the nuclear corrections 
enhance the $\pi^\pm$ channel and decrease the $\pi^0$ channel.\\

\noindent{\bf\underline{Example 2}}:
We consider next a realistic case in the Gargamelle experiment where they 
measured the production of pions in a enriched Freon target.
The average energy of the neutrino beam was low, in the 
$\Delta$-resonance region.
They observed the ratio \cite{musset}: 
\begin{eqnarray*}
\frac{\pi^+}{\pi^0}= 2.3 \pm 0.9.
\end{eqnarray*}
The cross sections for $E_\nu = 2.0 \, {\rm GeV}$ 
in units of $10^{-38} {\rm cm}^2$ are
\begin{eqnarray*}
\sigma(\nu p\rightarrow \mu^- p \pi^+) &=& 0.70\pm 0.10\\ 
\sigma(\nu n\rightarrow \mu^- p \pi^0) &=& 0.20\pm 0.05\\ 
\sigma(\nu n\rightarrow \mu^- n \pi^+) &=& 0.20\pm 0.07.\\ 
\end{eqnarray*}
Consequently the ratio is $\big(\frac{\pi^+}{\pi^0}\big)= 0.45$, far away 
from the result of Gargamelle. Including nuclear corrections brings 
the ratio in the proximity of the experimental number.
The pion yields after nuclear corrections are given by
\begin{eqnarray*}
{\rm d}\sigma(Br, \pi^+) &=& A\times(0.78\times0.90+0.16\times 0.20)\\
                   &=& A\times 0.74\\
{\rm d}\sigma(Br, \pi^0) &=& A\times(0.16\times0.90+0.68\times 0.20)\\ 
                   &=& A\times 0.28\\
\end{eqnarray*}
in units of $10^{-38} {\rm cm}^2$ and $Br$ denoting 
the nuclear content of the target. The ratio now is
\begin{eqnarray*}
\big(\frac{\pi^+}{\pi^0}\big)_{nucl. corr.} = \frac{0.74}{0.28} = 2.64 
\end{eqnarray*}
in good agreement with the Gargamelle measurement.\\

There are several other cases where comparisons were reported. 
These examples \cite{kluttig,merenyi} 
include absorption and charge exchange effects 
and find good agreement. 
It will be interesting to summarize these comparisons 
and to include the results of electroproduction experiments
in order to find out how well the model works.
There is also another model \cite{singh} 
where nuclear effects are 
incorporated in the scattering from a nuclear potential.
In this model particle absorption is included but 
there are no estimates of charge exchange effects.   

\subsection{Results}
\begin{figure}[htb]
\vspace{-1.cm}
\hspace{-1.cm}
\includegraphics[angle = 0,width =18pc,height = 15pc]{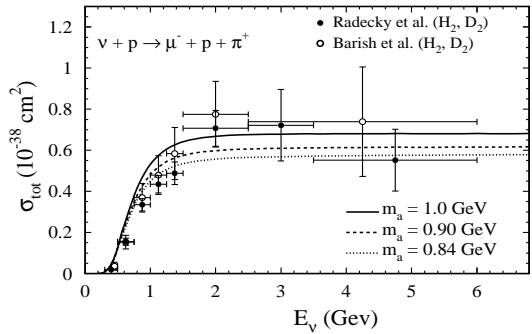}
\vspace{-1.8cm}
\caption{The CC cross sections for RES}
\label{fig4}
\end{figure}

\begin{figure}[htb]
\vspace{-1.cm}
\hspace{-1.cm}
\includegraphics[angle = 0,width =18pc,height = 15pc]{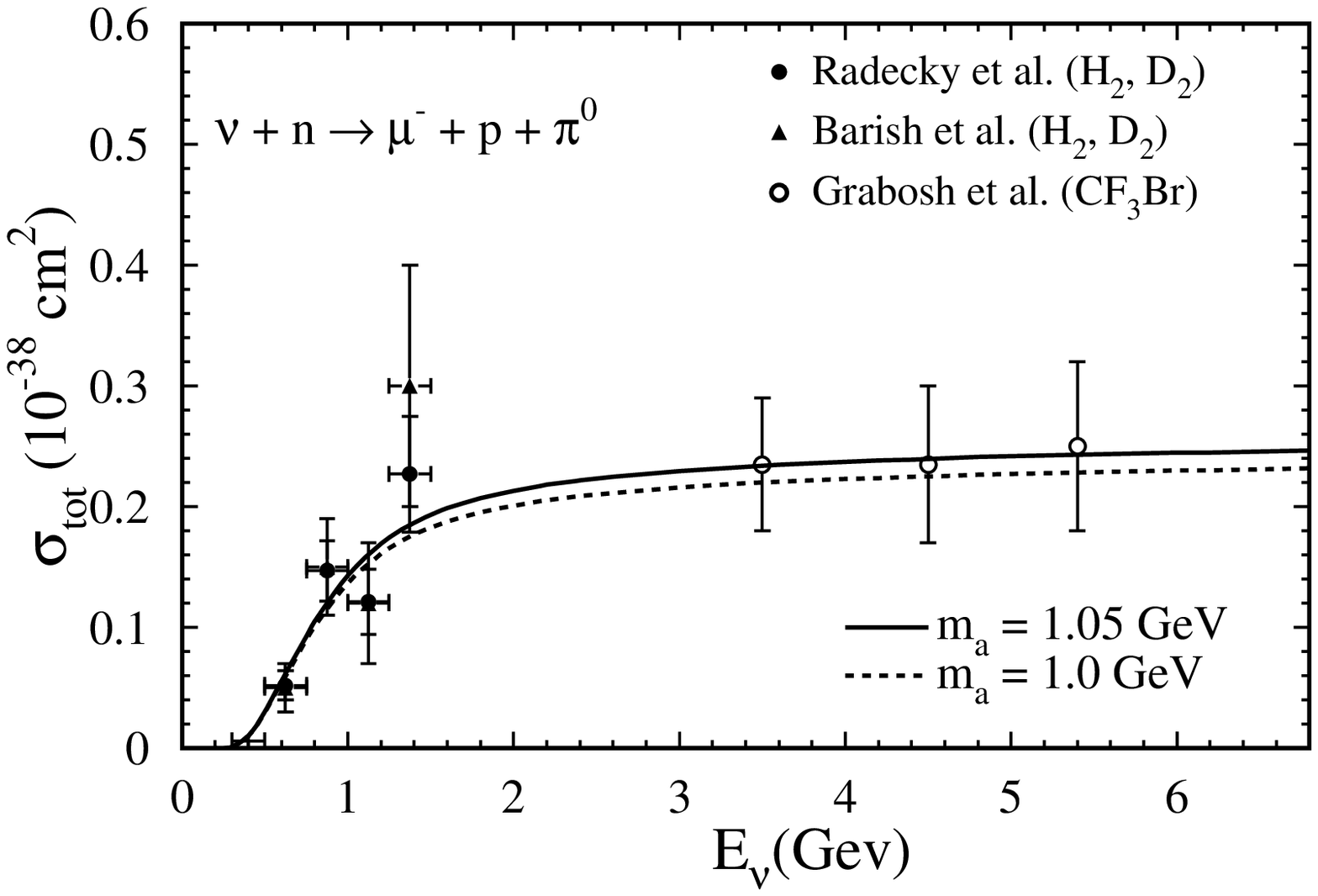}
\vspace{-1.8cm}
\caption{The CC cross sections for RES}
\label{fig5}
\end{figure}

\begin{figure}[htb]
\vspace{-0.7cm}
\hspace{-1.cm}
\includegraphics[angle = 0,width =18pc,height = 15pc]{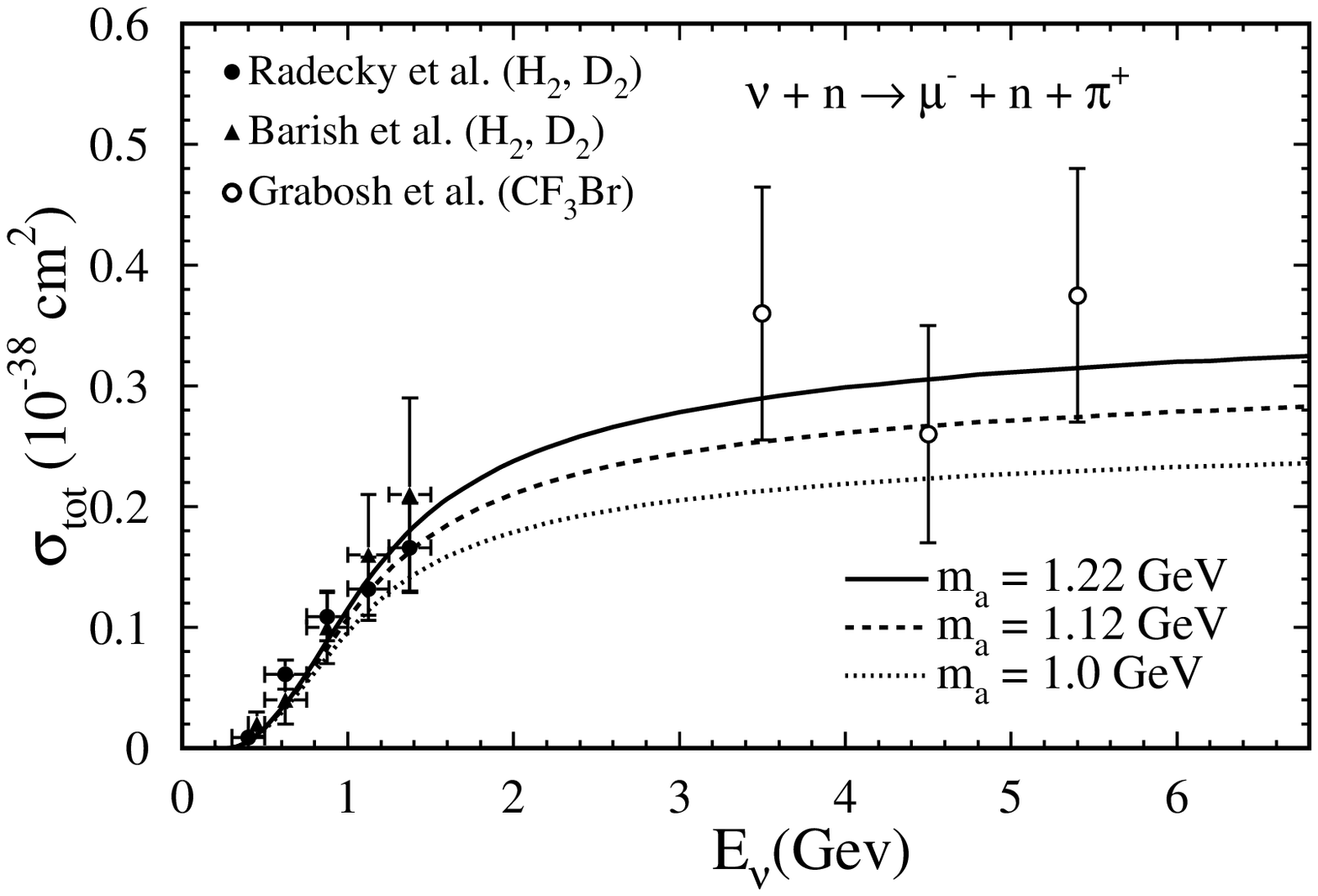}
\vspace{-1.8cm}
\caption{The CC cross sections for RES}
\label{fig6}
\end{figure}

\vspace{0.4cm}
We mentioned already the comparisons in Figures~(\ref{fig4})--(\ref{fig6}) 
where the integrated cross sections for the production of resonances 
$\Delta\, (1232), \,P_{11}\, (1440)$ and $S_{11}\, (1535)$ are compared with 
the available data. The agreement between theory and experiments is good in
the low energy regions where the experimental errors are small. 
At higher energies the errors are large and improvements are expected from 
the new experiments.

\vspace{-0.5cm}

In Figure~(\ref{fig7}), 
I show the cross sections for two neutral current reactions 
with the axial vector mass $m_a = 1\, {\rm GeV}$.
In addition we calculated the differential cross sections
$\frac{d\sigma}{d\,E_{\pi}}$. 
In Fig.~(\ref{fig8}) we show the spectrum from $\pi^+$'s on 
an Oxygen target and for a neutrino energy $E_\nu = 1.5 \,GeV$.
The dotted curve is without nuclear corrections, 
the broken line includes Pauli blocking. 

\begin{figure}[htb]
\vspace{-1.cm}
\hspace{-1.cm}
\includegraphics[angle = 0,width =18pc,height = 15pc]{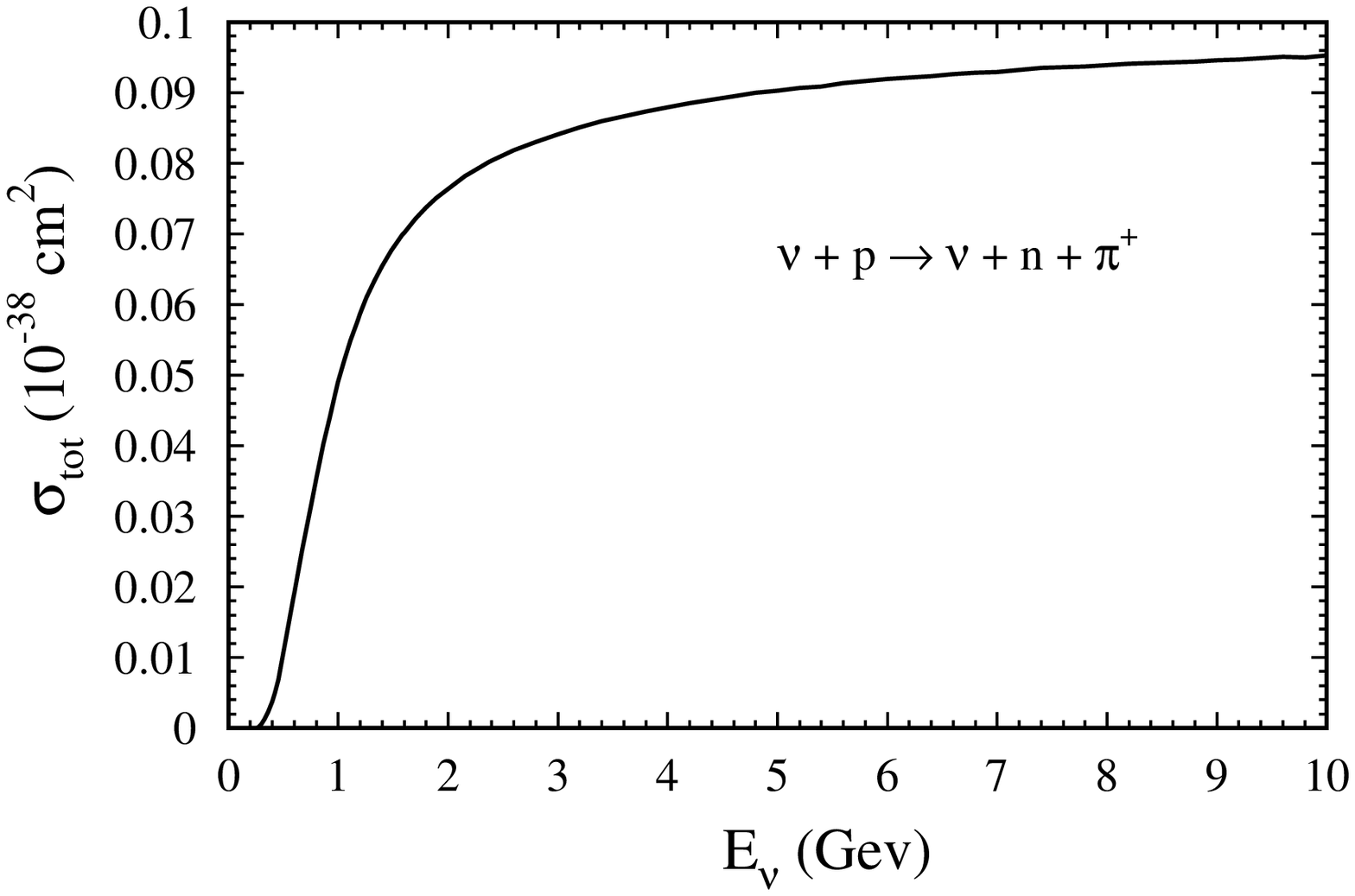}
\end{figure}
\begin{figure}[htb]
\vspace{-2.cm}
\hspace{-1.cm}
\includegraphics[angle = 0,width =18pc,height = 15pc]{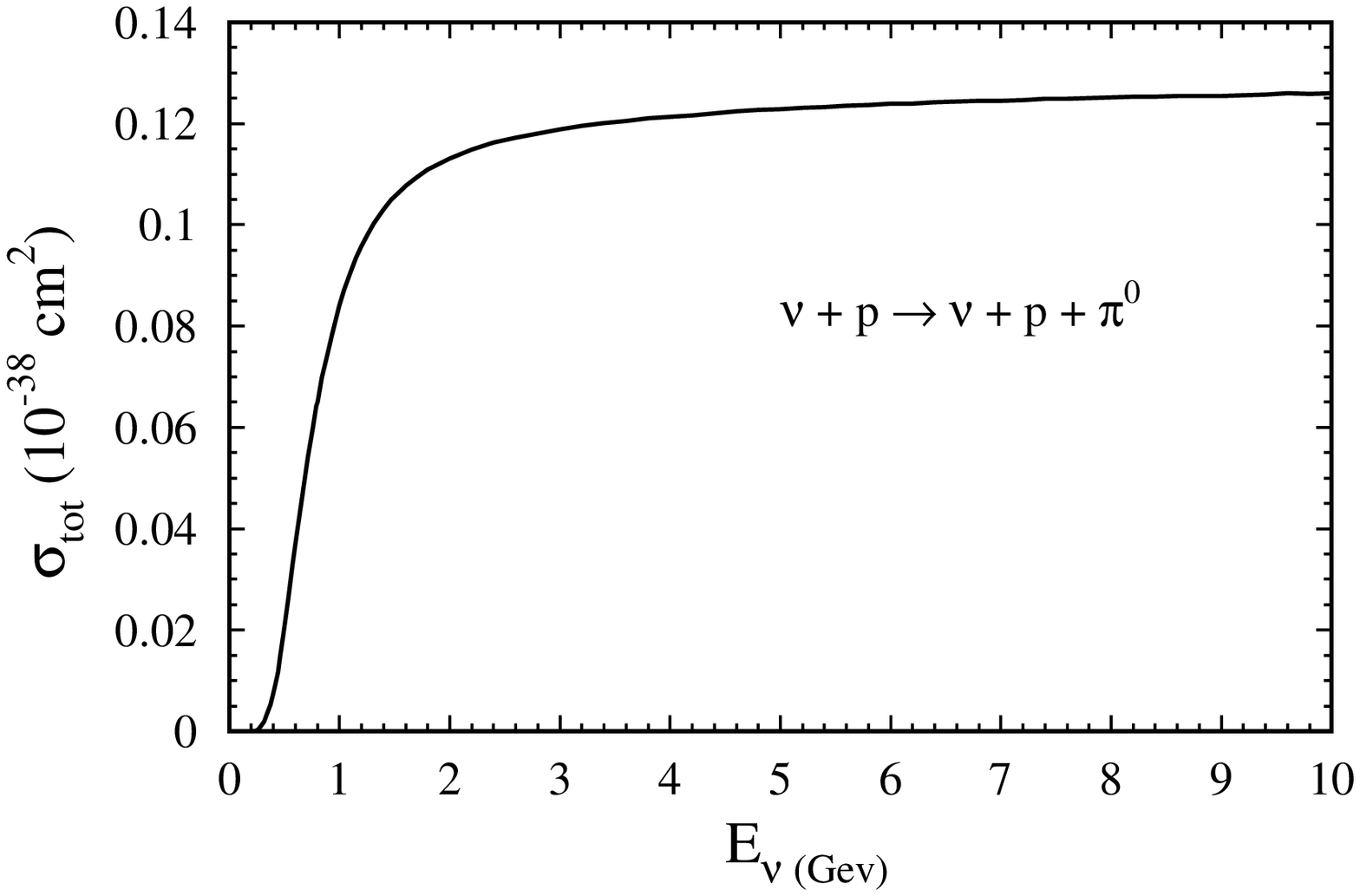}
\vspace{-1.5cm}
\hspace{-1.5cm}
\caption{The NC cross sections for resonance production with 
$m_a=1\,{\rm GeV}$}
\label{fig7}
\end{figure}
\begin{figure}[htb]
\vspace{-1.2cm}
\hspace{-1.cm}
\includegraphics[angle = 0,width =18pc,height = 14pc]{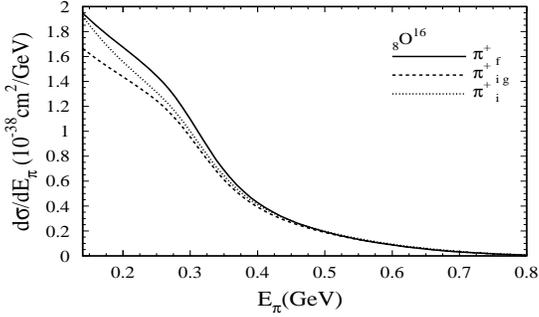}
\vspace{-1.5cm}
\caption{Pion energy distribution for $\pi^+$ on 
oxygen target \protect\cite{eap}}
\label{fig8}
\end{figure}

\begin{figure}[htb]
\vspace{-1.5cm}
\hspace{-1.cm}
\includegraphics[angle = 0,width =18pc,height = 14pc]{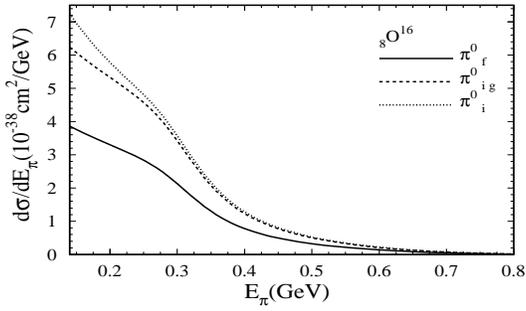}
\vspace{-1.5cm}
\hspace{-2.5cm}
\caption{Pion energy distribution for $\pi^0$ 
on oxygen target \protect\cite{eap}}
\label{fig9}
\end{figure}

Finally the solid curve includes absorption and charge-exchange corrections.
The curves in Figure~(\ref{fig9}) are the same as in Figure~(\ref{fig8}), 
but now for the production of $\pi^0$'s. 
We notice that the nuclear corrections are small for $\pi^+$ production 
and substantial for $\pi^0$'s.  

Now, since the neutrino beams peak at very low 
energy $\langle E_{\nu}\rangle=1.5$ GeV, the single
pion channels play an important role and should be
studied in the experiments.  In fact the experiments
should try to confirm or disprove the theoretical
calculations.

\section{Benchmarks}

\begin{figure}[htb]
\vspace{-1.2cm}
\hspace{-1.cm}
\includegraphics[angle = 0,width =18pc,height = 15pc]{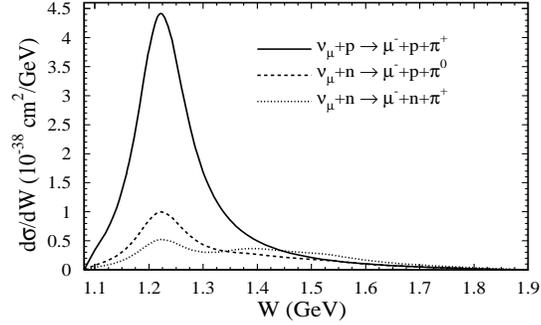}
\vspace{-2.cm}
\caption{The invariant mass distribution}
\label{fig10}
\end{figure}

It is interesting and important to provide some benchmark features 
of the calculations. For resonance production the invariant mass 
distribution of the final hadrons provide an outstanding feature.
In Figure~(\ref{fig10}), 
I show the mass distribution ${\rm d}\sigma/{\rm d} W$
for the dominant reactions and for $E_\nu = 1.0\, {\rm GeV}$ 
and $W$ is the pion-nucleon invariant mass. 
The cross sections were integrated over all values of $Q^2$, which appear 
in the form factors.

As a second benchmark I present several relations which are model independent.
The cross sections are frequently calculated on proton and neutron targets. 
The experiments, however, are carried out in nuclear targets, 
where as, we described, nuclear corrections are significant. 

During the conference, 
it became evident to me that we know so many properties of neutrino 
interactions that several ratios of neutral to charged current cross sections
can be determined without making the assumption of single nucleon interactions.
Instead we consider the scattering of the current from the 
$\underline{complete\, nucleus}$ leading to a final pion and a recoiling 
hadronic system (perhaps a recoiling new nucleus) 
with isospin $I = 0, 1$ or $2$.
Using conventional methods we can derive several relations, which I describe.

We define the cross sections on isoscalar targets
\begin{eqnarray*}
\sigma_-^{ch} &=& \sigma(\nu_\mu+(I=0)\rightarrow \mu^- +\pi^+ +x_1)\\
              &+& \sigma(\nu_\mu+(I=0)\rightarrow \mu^- +\pi^- +x_2)\\
\sigma_-^0    &=& \sigma(\nu_\mu+(I=0)\rightarrow \mu^- +\pi^0 +x_3)\\
\sigma_0^0    &=& \sigma(\nu_\mu+(I=0)\rightarrow \nu_\mu +\pi^0 +x_4)
\end{eqnarray*}
with $x_1,...,\, x_4$ the undetected hadronic states. 
We also define $\Sigma = \sigma_-^{ch}-\sigma_-^0 $ and for antineutrino
$\bar{\Sigma} = \sigma_+^{ch}-\sigma_+^0$.
Then the ratio 
\begin{eqnarray*}
R_0 = \frac{\sigma_0^0+\bar{\sigma}_0^0}{(\Sigma+\bar{\Sigma})} = 0.37
\end{eqnarray*}
is precisely determined with an uncertainty of at most $15\%$.

An alternative ratio is obtained by adding neutrino and antineutrino cross
section without observing the final hadronic states
\begin{eqnarray*}
R_1 = \frac{\sigma_0+\bar{\sigma_0}}{\sigma_-+\sigma_+}
\end{eqnarray*}
An estimate of the ratio in the high energy region gives 
the value $R_1 = 0.42$ with a small uncertainty of $5\%$.
A third ratio of this kind is the PW-relation \cite{wolfen}.

Other ratios and the method which leads to them are 
included in a recent article \cite{eap1}. 

It is important to use model independent relations 
or cross sections which automatically include nuclear effects. 
The reasons for this is to avoid misinterpretation of the phenomena.
Among the possible mistakes are the following:
\begin{enumerate}
\item[i)]The misidentification of charge pions as muons may interpret
hadronic mixing as neutrino
mixing.
\item[ii)] The mixing of $\pi^0$'s with charged pions
modifies the pion yield.  This influences
the estimate of how many $\pi^0$'s are produced by
neutral currents and may lead to the interpretation
that the missing $\pi^0$'s are explained as an oscillation
of active to sterile neutrinos.
\end{enumerate}

\begin{figure}[htb]
\vspace{-0.8cm}
\hspace{-1.cm}
\includegraphics[angle = 0,width =18pc,height = 15pc]{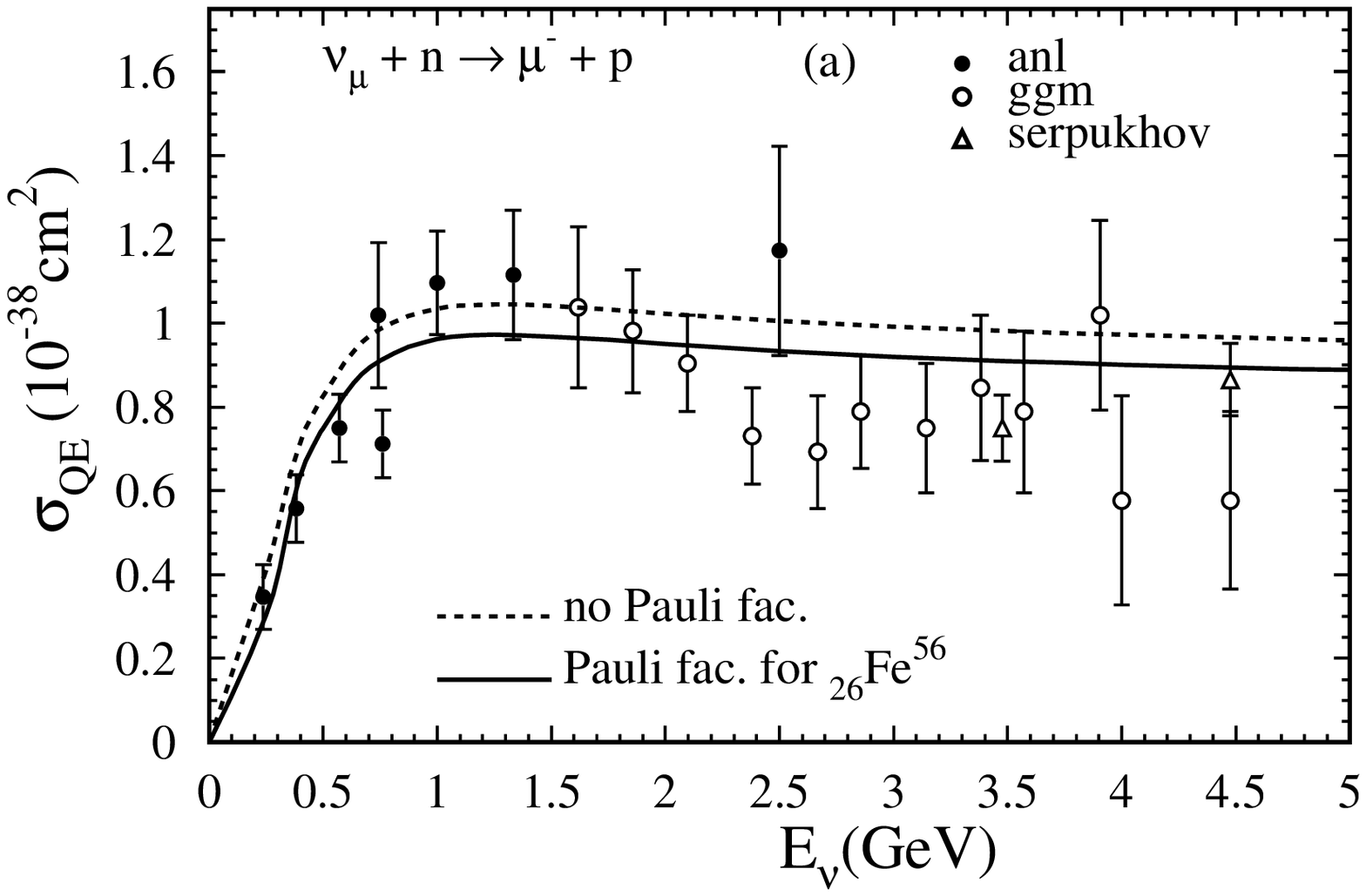}
\end{figure}
\begin{figure}[htb]
\vspace{-2.cm}
\hspace{-1.cm}
\includegraphics[angle = 0,width =18pc,height = 15pc]{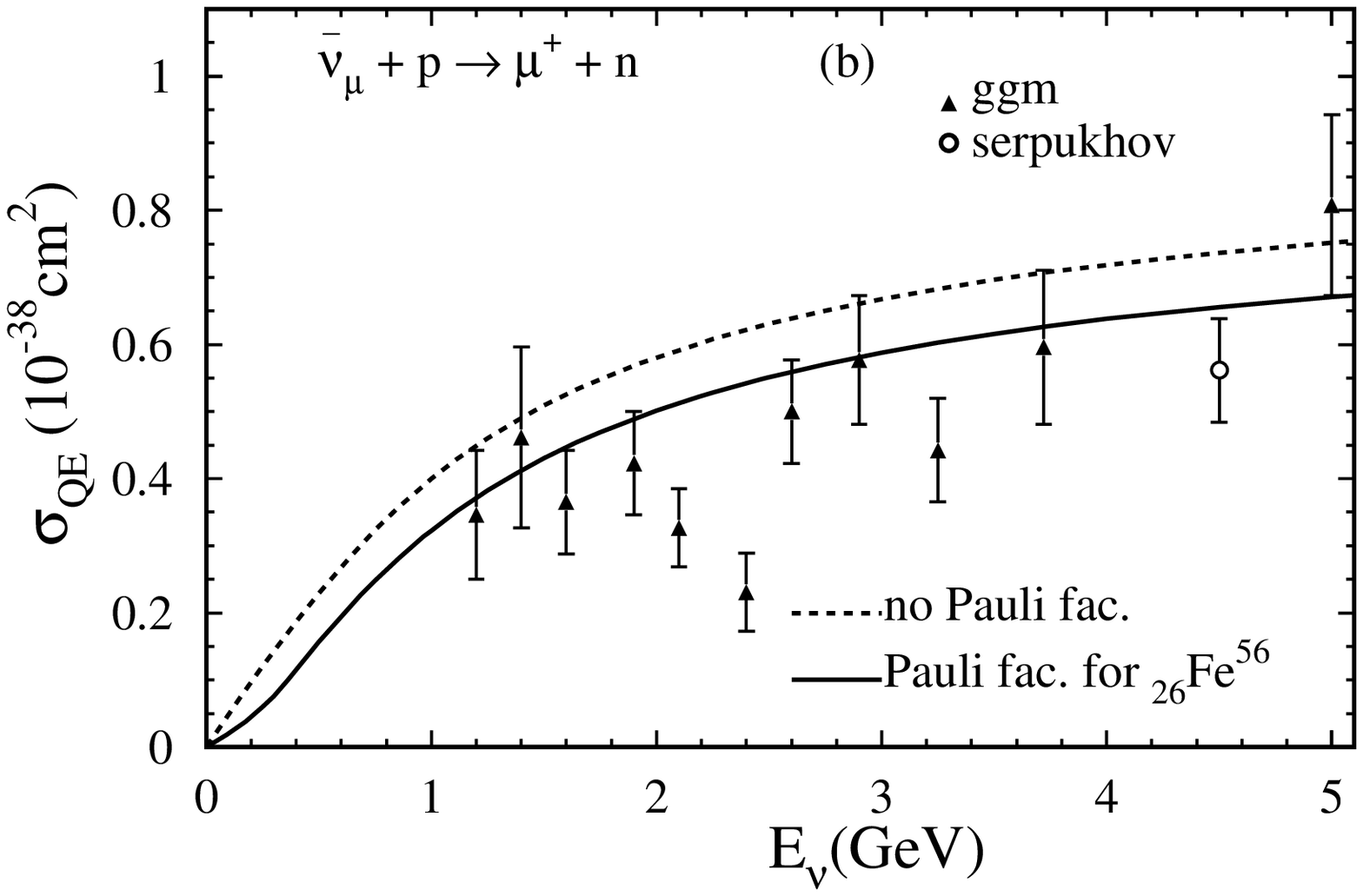}
\vspace{-1.5cm}
\hspace{-1.cm}
\caption{The CC cross sections of QE for neutrino and antineutrino process}
\label{fig11}
\end{figure}
\section{Quasi--elastic Scattering}

The reactions
\begin{eqnarray}
\nu +n      &\rightarrow& \mu^- + p \nonumber\\
\bar{\nu}+p &\rightarrow& \mu^+ +n \nonumber
\end{eqnarray}
have been calculated in terms of form factors.  
The quasi--elastic cross sections reach constant value at 
$E_\nu\geq 1\,{\rm GeV}$ 
and have been used for determining the fluxes of 
neutrino or antineutrino beams.
There are also measurements of these reactions. 

We show in Figure~(\ref{fig11}) the experimental data together
with theoretical calculations.
We show two curves with the dotted curve representing the scattering on free
neutrons and protons, while the solid curves include the reduction introduced 
by Pauli blocking \cite{yuji}.
The nuclear corrections are small, i.e. of the order of $10\%$.

\section{Ionization of Atoms by Neutrinos \cite{gounaris}}

As a last topic, 
I will discuss a very low energy reaction 
where a beam of $\bar{\nu}_e$ with an energy 
$E_{\bar{\nu}}\approx 12 \,{\rm keV}$ 
is produced in tritium decays
\begin{eqnarray*}
^3H \rightarrow ^3He + e^- + \bar{\nu}_e.
\end{eqnarray*}
The antineutrinos have very low energies 
and the oscillation length is greatly reduced.
For the presently favored solution (Large Mixing Angle solution) 
of the solar neutrino problem  the oscillation length is of the order of 
a few hundred meters or less.
We calculated the ionization cross section for H, He 
and  Ne atoms and found that 
$\sigma_{ionization}^{\bar{\nu}}/Z \approx$ (a few)$\times 10^{-47} 
{\rm cm}^2$.
Using realistic wave functions we found that the cross section for 
$E_{\bar{\nu}}>10 \,{\rm keV}$ is the incoherent sum 
of contributions from individual electrons. 
The cross section is small but there is interest 
for measuring the ionization with an intensive tritium source \cite{bouchez}.
Very low energy neutrino interactions are a terra incognita and may lead to 
surprises.

\section{Summary}

I presented cross sections for several neutrino induced 
reactions from very low energies up to the GeV region. 
Most of the calculations are reliable and in the cases where experimental 
data are available the agreement is rather good.
In the threshold regions, there are uncertainties 
and detailed studies will show how important they are when integrated over
the neutrino spectrum.
Additional corrections appear for nuclear targets. 
For quasi--elastic and deep inelastic scattering 
and for the energies we consider, nuclear corrections are small. 
We found large nuclear corrections for single pion production, coming 
from absorption and charge exchange interactions of the pions.
We studied a few channels in the ANP--Model where the corrections are 
substantial and modify the initial neutrino--nucleon 
interactions in right direction.

In order to avoid these uncertainties, 
I considered ratios of neutral to charged current cross sections where
the largest contribution is determined by symmetry considerations and 
smaller terms are estimated from data and/or theoretical calculations. 
For the LBL experiments the yields for the 
ratios should increase with the distance, 
because the $\tau$--neutrinos can not contribute to the charged current 
cross sections, since their energy is below the $\tau$--lepton threshold.

We hope that these calculations will be useful for the analysis 
and interpretation of the new experiments, especially those 
looking for oscillation phenomena. \\\\

\noindent{\bf Acknowledgments}\\

I wish to thank Dr. J.-Y. Yu for helping me in the 
preparation of this article.

\end{document}